\documentclass[osajnl,twocolumn,showpacs,superscriptaddress,10pt]{revtex4-1}
\usepackage{amsmath,amssymb,graphicx}
\begin{document}

\title{Soliton regulation in microcavities induced by fundamental--second-harmonic mode coupling}

\author{Xiaoxiao Xue}\email{Corresponding author: xuexx@tsinghua.edu.cn}
\author{Xiaoping Zheng}
\author{Bingkun Zhou}
\affiliation{Department of Electronic Engineering, Tsinghua University, Beijing 100084, China}

\begin{abstract}Microcomb generation with simultaneous $\chi^{(2)}$ and $\chi^{(3)}$ nonlinearities brings new possibilities for ultra-broadband and potentially self-referenced integrated comb sources. However, the evolution of the intracavity field involving multiple nonlinear processes shows complex dynamics that is still poorly understood. Here we report on strong soliton regulation induced by fundamental--second-harmonic (FD-SH) mode coupling. The formation of solitons from chaos is extensively investigated based on coupled Lugiato--Lefever equations. The soliton generation shows more deterministic behaviors in the presence of FD-SH mode interaction, in sharp contrast to the usual cases where the soliton number and relative locations are stochastic. Deterministic single soliton transition, soliton binding and prohibition are observed, depending on the phase matching condition and coupling coefficient between the fundamental and second-harmonic waves. Our finding provides important new insights into the soliton dynamics in microcavities with simultaneous $\chi^{(2)}$ and $\chi^{(3)}$ nonlinearities, and can be immediate guidance for broadband soliton comb generation with such platforms.
\end{abstract}


\maketitle 

\section{Introduction}

Microresonator based optical frequency comb (microcomb) generation is very promising to achieve highly compact, chip-scale integrated broadband comb sources \cite{ref1,ref2}. Many applications can potentially benefit from this revolutionary technology, such as high-speed optical communications \cite{ref3,ref4,ref5}, optical clocks \cite{ref6,ref7,ref8}, dual-comb spectroscopy \cite{ref9,ref10}, fast detecting lidar \cite{ref11,ref12}, THz generation \cite{ref13}, and microwave photonic signal generation and processing \cite{ref14,ref15,ref16,ref17,ref18,ref19,ref20}. Microcomb generation relies on Kerr effect which is a $\chi^{(3)}$ nonlinearity. Mode-locked comb states are generally related to formation of cavity solitons due to a balanced interplay between dispersion and Kerr nonlinearity \cite{ref21,ref22}. Nevertheless, many microresonator platforms such as SiN, AlN, AlGaAs and diamond also possess $\chi^{(2)}$ nonlinearity \cite{ref23,ref24,ref25,ref26}. (Although bulk amorphous SiN is generally considered having no $\chi^{(2)}$ nonlinearity, SiN waveguides may do have due to surface effects or silicon nanocrystals that form in the fabrication process.) Kerr comb generation accompanied by second-harmonic generation has been reported before with such microresonators \cite{ref27,ref28}. Recently, microcomb generation through simultaneous $\chi^{(2)}$ and $\chi^{(3)}$ nonlinearities is attracting increasing attention \cite{ref29,ref30,ref31} because it provides a novel way to extend the comb spectral range and brings new possibilities of single-resonator comb self-referencing which might greatly facilitate on-chip optical clocks \cite{ref32}. Moreover, the evolution of the optical fields involving coupled $\chi^{(2)}$ and $\chi^{(3)}$ nonlinearities shows complex dynamics that is of high scientific interest.

Dissipative cavity solitons are of high interest in the field of microcomb generation because they correspond to broadband mode-locked comb states. Intense studies have been performed to investigate the cavity soliton dynamics involving different physical processes, such as Raman effect \cite{ref33}, spatial mode coupling \cite{ref34,ref35,ref36}, and external seeding \cite{ref37}. Nevertheless, the impact of second-harmonic generation on the soliton dynamics has been little explored and remains poorly understood. Here we report on strong soliton regulation induced by fundamental--second-harmonic (FD--SH) mode coupling in microresonators with simultaneous $\chi^{(2)}$ and $\chi^{(3)}$ nonlinearities. We show through simulations based on coupled Lugiato--Lefever (L-L) equations that deterministic single soliton transition and soliton binding may be achieved depending on the phase matching condition and the coupling coefficient between the fundamental and second-harmonic waves. We also find conditions when stable soliton states are prohibited by strong FD-SH mode coupling thus it is difficult to achieve comb mode-locking. Our results provide important new insights into the soliton dynamics in microresonators with simultaneous $\chi^{(2)}$ and $\chi^{(3)}$ nonlinearities, and suggest that FD-SH mode coupling may be employed as a new way for soliton manipulation in microresonators.

\section{Theoretical Model}

The microcomb generation scheme we investigate is illustrated in Fig. 1(a). A microresonator with simultaneous $\chi^{(2)}$ and $\chi^{(3)}$ nonlinearities is pumped by a continuous-wave (cw) single frequency laser. The microresonator has anomalous group velocity dispersion in the fundamental wavelength range. In the cavity, both fundamental and second-harmonic combs are generated. Note that frequency comb generation here is attributed to Kerr nonlinearity, other than cascaded $\chi^{(2)}$ process reported previously \cite{ref38,ref39}. We numerically investigate the evolution of the optical fields based on the following coupled mean-field L-L equations \cite{ref29}.
\begin{align}\label{eq:Coupled-LLE1}
  \frac{\partial E_1}{\partial z} & = \bigg [ -\alpha_1 - \mathrm{i}\delta_1 - \mathrm{i} \frac{k_1^{''}}{2} \frac{\partial^2}{\partial \tau^2} + \mathrm{i}\gamma_1 \left|E_1\right|^2 \nonumber \\
  &+ \mathrm{i}2\gamma_{12}\left|E_2\right|^2 \bigg]E_1 + \mathrm{i}\kappa E_2 E_1^* + \eta_1 E_{\mathrm{in}} \\
  \label{eq:Coupled_LLE2}
  \frac{\partial E_2}{\partial z} & = \bigg [ -\alpha_2 - \mathrm{i}2\delta_1 -\mathrm{i}\Delta k -\Delta
  k'\frac{\partial}{\partial\tau} - \mathrm{i} \frac{k_2^{''}}{2} \frac{\partial^2}{\partial \tau^2} \nonumber \\
  &+ \mathrm{i}\gamma_2 \left|E_2\right|^2
  + \mathrm{i}2\gamma_{21}\left|E_1\right|^2 \bigg]E_2 + \mathrm{i}\kappa^* E_1^2
\end{align}
Here, $E_1$ and $E_2$ are fundamental and second-harmonic amplitudes scaled such that $|E_1|^2$ and $|E_2|^2$ represent the optical power; $z$ is propagation distance in the cavity; $\alpha_1$ and $\alpha_2$ are amplitude losses including the intrinsic loss and the coupling loss; $\delta_1={\delta_0}/{L}$ where $\delta_0$ is round-trip phase detuning between the pump laser and the cavity, given by $\delta_0=(\omega_0-\omega_{\mathrm{p}})t_{\mathrm{R}}$; $\omega_0$ and $\omega_{\mathrm{p}}$ are angular frequencies of the resonance and the pump laser, respectively; $t_{\mathrm{R}}$ round-trip time; $L$ round-trip length; $k_1^{''}=\mathrm{d}^2 k_{\mathrm{FD}}/\mathrm{d}\omega^2 \vert_{\omega=\omega_{\mathrm{p}}}$, $k_2^{''}=\mathrm{d}^2 k_{\mathrm{SH}}/\mathrm{d}\omega^2 \vert_{\omega=2\omega_{\mathrm{p}}}$ group velocity dispersion, where $k_{\mathrm{FD}}$ and $k_{\mathrm{SH}}$ are wave vectors of the fundamental and second-harmonic waves respectively (note that the fundamental and second-harmonic waves usually correspond to different spatial or polarization modes); $\Delta k=2k_{\mathrm{FD}}(\omega_{\mathrm{p}})-k_{\mathrm{SH}}(2\omega_{\mathrm{p}})$ wave vector mismatch; $\Delta k'=\mathrm{d}k_{\mathrm{SH}}/\mathrm{d}\omega\vert_{\omega=2\omega_{\mathrm{p}}}-\mathrm{d}k_{\mathrm{FD}}/\mathrm{d}\omega\vert_{\omega=\omega_{\mathrm{p}}}$ group velocity mismatch; $\gamma_1$, $\gamma_2$ nonlinear coefficients of self-phase modulation; $\gamma_{12}$, $\gamma_{21}$ nonlinear coefficients of cross-phase modulation; $\kappa$ second-order coupling coefficient; $E_{\mathrm{in}}$ pump amplitude; $\eta_1=\sqrt{\theta_1}/L$ where $\theta_1$ is power coupling coefficient between the bus waveguide and the microcavity for the fundamental wave.

\begin{figure}[tbp]
\centerline{\includegraphics[width=0.8\columnwidth]{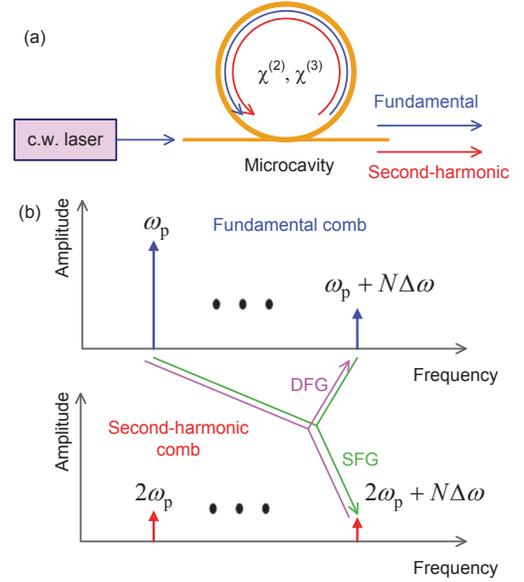}}
\caption{(a) Scheme of Kerr comb generation in a microcavity with simultaneous $\chi^{(2)}$ and $\chi^{(3)}$ nonlinearities. (b) Illustration of mode coupling between the fundamental and second-harmonic waves through sum (SFG) and difference (DFG) frequency generation.}
\end{figure}

It has been reported in Refs. \cite{ref29} and \cite{ref40} that the sideband mode of the fundamental wave $(\omega_{\mathrm{p}}+N\Delta\omega)$ may be strongly coupled to the sideband mode of the second-harmonic wave $(2\omega_{\mathrm{p}}+N\Delta\omega)$ through sum and difference frequency generation (illustrated in Fig. 1(b)), i.e. $\omega_{\mathrm{p}}+(\omega_{\mathrm{p}}+N\Delta\omega)\rightarrow (2\omega_{\mathrm{p}}+N\Delta\omega)$ and $(2\omega_{\mathrm{p}}+N\Delta\omega)-\omega_{\mathrm{p}}\rightarrow (\omega_{\mathrm{p}}+N\Delta\omega) $ where $N$ is the relative mode number and $\Delta\omega=2\pi FSR$. We consider the case when the fundamental field is composed of stable solitons in the cavity. To estimate the phase matching condition for the sum and difference frequency conversion, we assume the power of the second-harmonic wave is much weaker than that of the fundamental wave, so that the soliton state is barely affected by the FD-SH mode coupling. For stable soliton states, the dispersion is exactly balanced by the Kerr nonlinearity. Therefore, the round-trip phase detuning for all the fundamental modes is zero. Note that for the pump mode, the external pump injection should be taken into account to yield the overall zero phase detuning. The round-trip phase mismatch for the sum frequency conversion $\omega_{\mathrm{p}}+(\omega_{\mathrm{p}}+N\Delta\omega)\rightarrow (2\omega_{\mathrm{p}}+N\Delta\omega)$ can then be written as
\begin{align}
\Delta\phi &= \phi_{\mathrm{SH}}\left( 2\omega_{\mathrm{p}}+N\Delta\omega\right) - \phi_{\mathrm{FD}}\left( \omega_{\mathrm{p}} \right) \nonumber \\
& - \phi_{\mathrm{FD}}\left( \omega_{\mathrm{p}}+N\Delta\omega\right) \nonumber \\
&= -\Delta kL+\Delta k' LN\Delta\omega + \frac{k_2^{''}L}{2}\left(N\Delta\omega\right)^2 \nonumber \\
&+ 2\gamma_{21}\left<|E_1|^2 \right>L - 2\delta_0
\end{align}
where $\phi_{\mathrm{FD}}(\omega_{\mathrm{p}})=0$, $\phi_{\mathrm{FD}}(\omega_{\mathrm{p}}+N\Delta\omega)=0$ for the pump and sideband modes of the fundamental soliton; $\left<|E_1|^2\right>$ represents the mean power of the fundamental wave. Here we omit the self-phase modulation of the second-harmonic mode since its power is assumed very weak.

\section{Results and Discussion}

Numerical simulations are performed to validate Eq.~(3). We use typical parameters for a SiN microring resonator as follows: $FSR=200\ \mathrm{GHz}$, $L=730\ \mathrm{\mu m}$, $\alpha_1=3.038\times10^{-3}\ \mathrm{m}^{-1}$, $\alpha_2=6.076\times10^{-3}\ \mathrm{m}^{-1}$, $k_1^{''}=-100\ \mathrm{ps^2/km}$, $k_2^{''}=200\ \mathrm{ps^2/km}$, $\Delta k'=7.61\times10^{-10}\ \mathrm{s\cdot m}^{-1}$, $\gamma_1=0.8\ \mathrm{m}^{-1}\mathrm{W}^{-1}$, $\gamma_2=2.1\ \mathrm{m}^{-1}\mathrm{W}^-1$, $\gamma_{12}=0.6\ \mathrm{m}^{-1}\mathrm{W}^{-1}$, $\gamma_{21}=1.2\ \mathrm{m}^{-1}\mathrm{W}^{-1}$, $\eta_1=75.51\ \mathrm{m}^{-1}$, $P_{\mathrm{in}}=|E_{\mathrm{in}}|^2=0.2\ \mathrm{W}$, $\kappa=2$, $\delta_0=0.03$. The initial fundamental field in the cavity is a single soliton given by \cite{ref21}
\begin{align}
E_1\vert_{z=0} &= E_{10} \nonumber \\ &+e^{\mathrm{i}\phi_0}\sqrt{\frac{2\delta_0}{\gamma_1L}} \mathrm{sech}\left[\sqrt{\frac{2\delta_0}{|k_1^{''}|L}}\left(\tau-\frac{t_{\mathrm{R}}}{2}\right)\right]
\end{align}
where $E_{10}$ is the lower-branch cw solution of Eq.~(1) with $E_2=0$; the phase term $\phi_0$ is given by
\begin{align}
\phi_0 = \mathrm{acos}\left(\frac{1}{\pi}\sqrt{\frac{8\delta_0\alpha_1^2}{P_{\mathrm{in}}\gamma_1\eta_1^2L}}\right)
\end{align}

The initial second-harmonic field is zero, i.e. $E_2\vert_{z=0}=0$. Equations (1) and (2) are then integrated with the split-step Fourier method until $E_1$ and $E_2$ reach stable states. We freely tune the wave vector mismatch term ($\Delta k$) to achieve phase matching for the 5th mode (i.e. $N=5$). By setting Eq. (3) to zero, the round-trip phase detuning induced by wave vector mismatch ($\Delta kL$) is calculated to be $-3.544$. We thus run the simulations with $\Delta kL$ scanned in a small range around $-3.544$. Figure 2(a) shows the power of the 5th second-harmonic mode versus $\Delta kL$. The peak position agrees with the theoretical prediction quite well. Figure 2(b) shows the fundamental and second-harmonic spectra when $\Delta kL=-3.544$ (i.e. nearly perfect phase matching). A strong peak at the 5th second-harmonic mode can be clearly observed ($>40\ \mathrm{dB}$ higher than the other modes). Figure 2(c) shows the time-domain waveforms. The second-harmonic wave shows as a step clamped with the fundamental soliton. Figure 2(d) shows the fundamental and second-harmonic spectra when $\Delta kL=-3.5$ (i.e. not perfect phase matching). The power of the 5th second-harmonic mode is much weaker ($\sim17\ \mathrm{dB}$ lower) than that in Fig. 2(b). Figure 2(e) shows the time-domain waveforms in this case.

\begin{figure}[tbp]
\centerline{\includegraphics[width=\columnwidth]{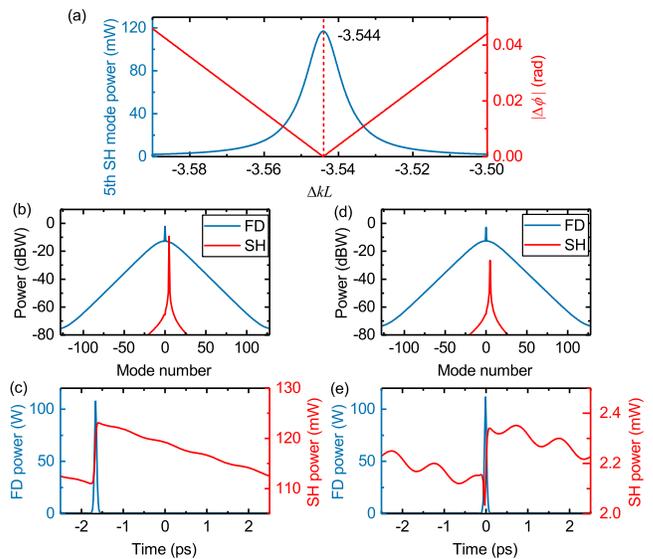}}
\caption{(a) The 5th second-harmonic mode power and the absolute phase mismatch ($|\Delta\phi|$) versus $\Delta kL$. The vertical dash line indicates the calculated $\Delta kL$ for perfect phase matching. (b) Spectra and (c) waveforms of the fundamental (FD) and second-harmonic (SH) waves when $\Delta kL=-3.544$ (perfect phase matching). (d) Spectra and (e) waveforms when $\Delta kL=-3.5$ (not perfect phase matching).}
\end{figure}

\begin{figure}[tbp]
\centerline{\includegraphics[width=\columnwidth]{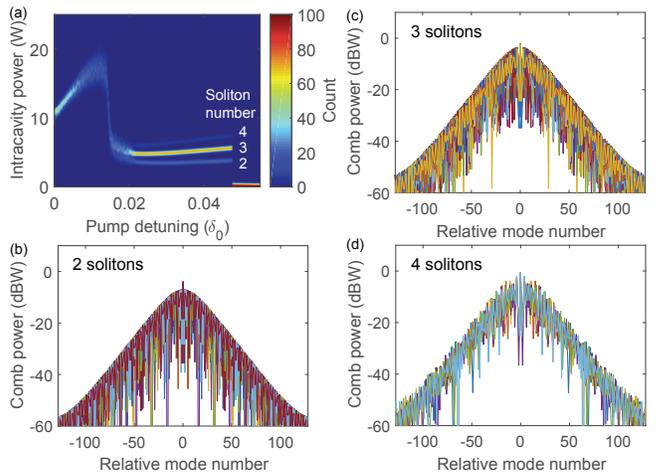}}
\caption{Soliton transitions in a cavity with only third-order Kerr nonlinearity ($\kappa=0$). (a) Intracavity power versus pump detuning. 100 traces are overlaid. The number of traces corresponding to 2, 3, and 4 solitons is 21, 73, and 6, respectively. (b), (c), (d) Overlaid comb spectra for 2, 3, and 4 solitons.  }
\end{figure}

We then investigate the impact of FD-SH mode coupling on the soliton formation dynamics. For comparison, we first run simulations by assuming no second-order nonlinearity (i.e. $\kappa=0$). In a traditional Kerr comb generation scheme, the pump laser is tuned from shorter wavelength to longer wavelength into the microresonator resonance. The intracavity field goes through a chaotic state before solitons are formed. Therefore, soliton transitions are usually stochastic, namely the number of solitons and their relative locations are random from test to test \cite{ref21}. In our simulations, the round-trip pump detuning ($\delta_0$) is scanned from 0 to 0.055 with a small step of 0.0001. For each step, the coupled equations are integrated for 500 round trips. The simulation parameters are as mentioned above. A total of 100 simulation tests are performed. The traces of the intracavity power versus the pump detuning are overlaid in Fig. 3(a). The intracavity field transitions from chaos to solitons when $\delta_0>0.02$. Discrete power steps corresponding to different number of solitons can be clearly observed. The occurrence count of 2-, 3-, and 4-soliton transitions is 21, 73, and 6, respectively. No single soliton state is observed. Figures 3(b), 3(c), and 3(d) show the overlaid comb spectra for 2, 3, and 4 solitons. The spectral envelope changes randomly from test to test, suggesting stochastic soliton locations in the cavity.

\begin{figure}[htbp]
\centerline{\includegraphics[width=\columnwidth]{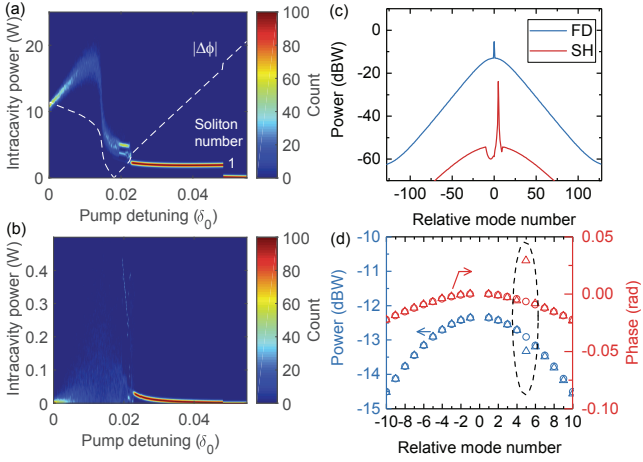}}
\caption{Deterministic single soliton formation assisted by FD-SH mode coupling when $\kappa=3$ and $\Delta kL=-3.515$. (a) Intracavity fundamental power versus pump detuning. 100 traces are overlaid. The white dash line shows the absolute phase mismatch calculated according to Eq. (3). The full vertical scale corresponds to $0\sim0.1\ \mathrm{rad}$. (b) Intracavity second-harmonic power versus pump detuning. (c) Spectra of the fundamental (FD) and second-harmonic (SH) waves when $\delta_0=0.025$. (d) Zoom-in spectra of the fundamental comb lines around the pump, showing weak amplitude and phase perturbations at the 5th mode (marked with a dashed circle). Circle ($\circ$): $\kappa=0$; triangle ($\triangle$): $\kappa=3$. }
\end{figure}

In the next, we run the simulations with $\kappa\neq0$ and tune the wave vector mismatch term ($\Delta k$) in Eq. (2). We find that the soliton formation dynamics can be drastically changed by the FD-SH mode coupling. Figures 4(a) and 4(b) show the power transition traces for the fundamental and second-harmonic waves when $\kappa=3$ and  $\Delta kL=-3.515$. Again, a total of 100 simulation tests are performed. Remarkably, in all the tests the fundamental field deterministically transitions to a single soliton state. The absolute value of phase mismatch calculated according to Eq. (3) is also plotted in Fig. 4(a). Nearly perfect phase matching is achieved around $\delta_0\sim0.02$. The formation of solitons is strongly affected by the FD-SH interaction, resulting in deterministic single soliton transition. Note that the phase mismatch for FD-SH mode coupling is not constant and dynamically changes with the pump detuning due to a direct contribution from the detuning (see Eq. (3)) and the cross-phase modulation from the fundamental wave.

\begin{figure}[tbp]
\centerline{\includegraphics[width=\columnwidth]{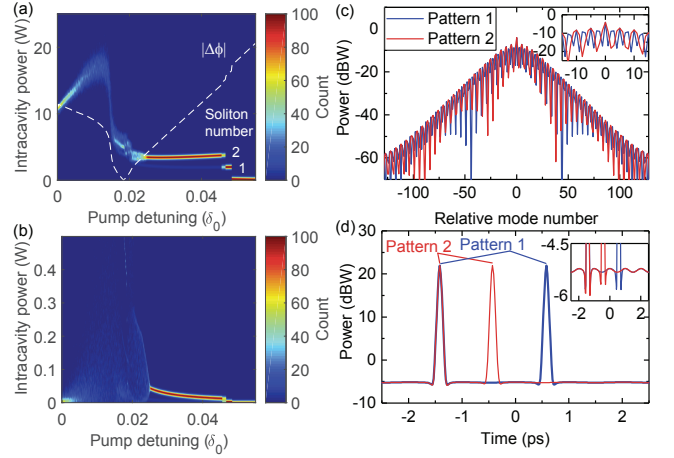}}
\caption{Soliton binding due to FD-SH mode coupling when $\kappa=3$ and $\Delta kL=-3.505$. (a) Intracavity fundamental power versus pump detuning. 100 traces are overlaid. The absolute phase mismatch calculated according to Eq. (3) is also plotted as in Fig. 4(a). (b) Intracavity second-harmonic power versus pump detuning. (c) Overlaid spectra of the fundamental comb when $\delta_0=0.045$. Two patterns can be observed. (d) Time-domain waveforms of the two patterns, showing that the soliton is trapped by the oscillations induced by FD-SH mode coupling. The time offset is adjusted such that one soliton is aligned for the two patterns.}
\end{figure}

Deterministic single soliton generation assisted by spatial mode coupling was reported recently and attributed to the interaction between solitons and dispersive wave \cite{ref36}. Interestingly, in our case of FD-SH mode coupling, the spectral variation related to dispersive wave may hardly be observed from the soliton spectrum which is plotted in Fig. 4(c) ($\delta_0=0.025$). In Fig. 4(d), we show zoom-in spectrum of the comb lines around the pump. The signature induced by FD-SH mode coupling now becomes distinguishable. The power variation of the 5th mode is only 0.42~dB, and the phase variation is 0.036~rad. Such weak perturbations are easily overlooked in experiments. Therefore, simultaneously monitoring both the fundamental and the second-harmonic spectra is highly desired to reveal the mode coupling. The perturbations of the 5th mode amplitude and phase correspond to background oscillations (i.e. dispersive wave) which may interact with the solitons, as will be shown in the next.

We find that the soliton transition is very sensitive to the phase matching condition of FD-SH mode coupling. Figures 5(a) and 5(b) shows the power transition traces of the fundamental and second-harmonic waves when $\kappa=3$ and $\Delta kL=-3.505$. 96 tests transition to 2 solitons while only 4 tests transition to single soliton. Figure 5(c) shows the overlaid spectra of the 2-soliton combs. In sharp contrast to the case with no FD-SH mode coupling (see Fig. 3(b)), the comb spectra here show two specific patterns. The occurrence count of pattern 1 and pattern 2 is 59 and 37, respectively. Figure 5(d) shows the time-domain waveforms for the two patterns. The inset shows a zoom-in plot of the background. It can be clearly observed that the solitons are trapped by the oscillations induced by FD-SH mode coupling. The perturbations to the 5th mode correspond to 5 oscillating periods in the time domain, dividing the cavity round trip to 5 slots. If we use 1 and 0 to represent whether or not there is a soliton in each slot, the two patterns can then be written as [10100] and [11000]. We attribute the soliton binding here to a similar mechanism that is induced by spatial mode coupling, higher-order dispersion, and external seeding \cite{ref41,ref42}.

We also find cases when soliton formation is prohibited by strong FD-SH interaction. Figure 6 shows the simulation results when $\kappa=6$ and $\Delta kL=-3.515$. All 100 simulations transition to a cw state from the chaotic state. No stable solitons are observed. This behavior is again similar to that induced by spatial mode coupling when strong coupling happens to the modes close to the pump \cite{ref34}. Our results suggest that FD-SH coupling should be carefully optimized to facilitate soliton comb generation in microresonators with simultaneous $\chi^{(2)}$ and $\chi^{(3)}$ nonlinearities.

\begin{figure}[tbp]
\centerline{\includegraphics[width=\columnwidth]{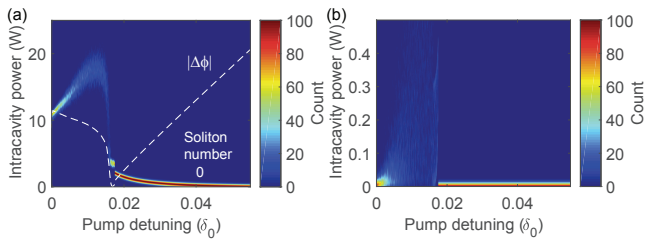}}
\caption{Soliton prohibition caused by strong FD-SH mode coupling when $\kappa=6$ and $\Delta kL=-3.515$. (a) Intracavity fundamental power versus pump detuning. 100 traces are overlaid. The absolute phase mismatch calculated according to Eq. (3) is also plotted as in Fig. 4(a). (b) Intracavity second-harmonic power versus pump detuning.}
\end{figure}

\section{Summary}

In summary, the impact of FD-SH mode coupling on soliton dynamics has been investigated through extensive simulations based on coupled L-L equations. The soliton formation is strongly regulated by FD-SH mode coupling. Deterministic single soliton transition, soliton binding and prohibition may be achieved, depending on the coupling coefficient and the phase matching condition. Our findings provide immediate guidance for soliton Kerr comb generation in microresonators with simultaneous $\chi^{(2)}$ and $\chi^{(3)}$ nonlinearities. Note that for simplicity, Ramman effect and higher-order dispersion are not considered in our current simulations. It has been known that Ramman effect and high-order dispersion may cause soliton self-frequency shift and spectral recoil \cite{ref33,ref35}. Both are expected to interact with FD-SH mode coupling and modify the phase matching condition. Another effect that is not considered is the thermo-optic effect which may also affect FD-SH phase matching if the thermo-optic coefficients are different in the fundamental and second-harmonic wavelength ranges. An advanced model including the higher-order effects will be worth investigating in the future.

{\bf Funding.} National Natural Science Foundation of China (61690191, 61690192, 61420106003); Beijing Natural Science Foundation (4172029).


\begin{thebibliography}{99}

\bibitem{ref1} T.~J.~Kippenberg, R.~Holzwarth, and S.~A.~Diddams, ``Microresonator-Based Optical Frequency Combs,'' Science \textbf{332}, 555--559 (2011).
\bibitem{ref2} A.~Pasquazi, M.~Peccianti, L.~Razzari, D.~J.~Moss, S.~Coen, M.~Erkintalo, Y.~K.~Chembo, T.~Hansson, S.~Wabnitz, P.~Del'Haye, X.~Xue, A.~M.~Weiner, and R.~Morandotti, ``Micro-combs: A novel generation of optical sources,'' Phys. Rep. \textbf{729}, 1--81 (2018).
\bibitem{ref3} P.~Marin-Palomo, J.~N.~Kemal, M.~Karpov, A.~Kordts, J.~Pfeifle, M.~H.~P.~Pfeiffer, P.~Trocha, S.~Wolf, V.~Brasch, M.~H.~Anderson, R.~Rosenberger, K.~Vijayan, W.~Freude, T.~J.~Kippenberg, and C.~Koos, ``Microresonator-based solitons for massively parallel coherent optical communications,'' Nature \textbf{546}, 274--279 (2017).
\bibitem{ref4} A.~F\"ul\"op, M.~Mazur, A.~Lorences-Riesgo, T.~A.~Eriksson, P.-H.~Wang, Y.~Xuan, D.~E.~Leaird, M.~Qi, P.~A.~Andrekson, A.~M.~Weiner, and V.~Torres-Company, ``Long-haul coherent communications using microresonator-based frequency combs,'' Opt. Exp. \textbf{25}, 26678--26688 (2017).
\bibitem{ref5} J.~Pfeifle, A.~Coillet, R.~Henriet, K.~Saleh, P.~Schindler, C.~Weimann, W.~Freude, I.~V.~Balakireva, L.~Larger, C.~Koos, and Y.~K.~Chembo, ``Optimally Coherent Kerr Combs Generated with Crystalline Whispering Gallery Mode Resonators for Ultrahigh Capacity Fiber Communications,'' Phys. Rev. Lett. \textbf{114}, 093902 (2015).
\bibitem{ref6} A.~A.~Savchenkov, D.~Eliyahu, W.~Liang, V.~S.~Ilchenko, J.~Byrd, A.~B.~Matsko, D.~Seidel, and L.~Maleki, ``Stabilization of a Kerr frequency comb oscillator,'' Opt. Lett. \textbf{38}, 2636--2639 (2013).
\bibitem{ref7} S.~B.~Papp, K.~Beha, P.~Del¡¯Haye, F.~Quinlan, H.~Lee, K.~J.~Vahala, and S.~A.~Diddams, ``Microresonator frequency comb optical clock,'' Optica \textbf{1}, 10--14 (2014).
\bibitem{ref8} V.~Brasch, E.~Lucas, J.~D.~Jost, M.~Geiselmann, and T.~J.~Kippenberg, ``Self-referenced photonic chip soliton Kerr frequency comb,'' Light Sci. Appl. \textbf{6}, e16202 (2017).
\bibitem{ref9} M.-G.~Suh, Q.-F.~Yang, K.~Y.~Yang, X.~Yi, and K.~J.~Vahala, ``Microresonator soliton dual-comb spectroscopy,'' Science \textbf{354}, 600--603 (2016).
\bibitem{ref10} A.~Dutt, C.~Joshi, X.~Ji, J.~Cardenas, Y.~Okawachi, K.~Luke, A.~L.~Gaeta, and M.~Lipson, ``On-chip dual-comb source for spectroscopy,'' Sci. Adv. \textbf{4}, e1701858 (2018).
\bibitem{ref11} P.~Trocha, M.~Karpov, D.~Ganin, M.~H.~P.~Pfeiffer, A.~Kordts, S.~Wolf, J.~Krockenberger, P.~Marin-Palomo, C.~Weimann, S.~Randel, W.~Freude, T.~J.~Kippenberg, and C.~Koos, ``Ultrafast optical ranging using microresonator soliton frequency combs,'' Science \textbf{359}, 887--891 (2018).
\bibitem{ref12} M.-G.~Suh and K.~J.~Vahala, ``Soliton microcomb range measurement,'' Science \textbf{359}, 884--887 (2018).
\bibitem{ref13} S.-W.~Huang, J.~Yang, S.-H.~Yang, M.~Yu, D.-L.~Kwong, T.~Zelevinsky, M.~Jarrahi, and C.~W.~Wong, ``Globally Stable Microresonator Turing Pattern Formation for Coherent High-Power THz Radiation On-Chip,'' Phys. Rev. X \textbf{7}, 041002 (2017).
\bibitem{ref14} X.~Xue and A.~M.~Weiner, ``Microwave photonics connected with microresonator frequency combs,'' Front. Optoelectron. \textbf{9}, 238--248 (2016).
\bibitem{ref15} W.~Liang, D.~Eliyahu, V.~S.~Ilchenko, A.~A.~Savchenkov, A.~B.~Matsko, D.~Seidel, and L.~Maleki, ``High spectral purity Kerr frequency comb radio frequency photonic oscillator,'' Nat. Comm. \textbf{6}, 7957 (2015).
\bibitem{ref16} S.~Diallo and Y.~K.~Chembo, ``Optimization of primary Kerr optical frequency combs for tunable microwave generation,'' Opt. Lett. \textbf{42}, 3522--3525 (2017).
\bibitem{ref17} X.~Xue, Y.~Xuan, H.-J.~Kim, J.~Wang, D.~E.~Leaird, M.~Qi, and A.~M.~Weiner, ``Programmable Single-Bandpass Photonic RF Filter Based on Kerr Comb from a Microring,'' J. Lightwave Technol. \textbf{32}, 3557--3565 (2014).
\bibitem{ref18} X.~Xue, Y.~Xuan, C.~Bao, S.~Li, X.~Zheng, B.~Zhou, M.~Qi, and A.~M.~Weiner, ``Microcomb-based true-time-delay network for microwave beamforming with arbitrary beam pattern control,'' J. Lightwave Technol. \textbf{36}, 2312--2321 (2018).
\bibitem{ref19} T.~G.~Nguyen, M.~Shoeiby, S.~T.~Chu, B.~E.~Little, R.~Morandotti, A.~Mitchell, and D.~J.~Moss, ``Integrated frequency comb source based Hilbert transformer for wideband microwave photonic phase analysis,'' Opt. Exp. \textbf{23}, 22087--22097 (2015).
\bibitem{ref20} X.~Xu, J.~Wu, M.~Shoeiby, T.~G.~Nguyen, S.~T.~Chu, B.~E.~Little, R.~Morandotti, A.~Mitchell, and D.~J.~Moss, ``Reconfigurable broadband microwave photonic intensity differentiator based on an integrated optical frequency comb source,'' APL Photon. \textbf{2}, 096104 (2017).
\bibitem{ref21} T.~Herr, V.~Brasch, J.~D.~Jost, C.~Y.~Wang, N.~M.~Kondratiev, M.~L.~Gorodetsky, and T.~J.~Kippenberg, ``Temporal solitons in optical microresonators,'' Nat. Photon. \textbf{8}, 145--152 (2014).
\bibitem{ref22} X.~Xue, Y.~Xuan, Y.~Liu, P.-H.~Wang, S.~Chen, J.~Wang, D.~E.~Leaird, M.~Qi, and A.~M.~Weiner, ``Mode-locked dark pulse Kerr combs in normal-dispersion microresonators,'' Nat. Photon. \textbf{9}, 594--600 (2015).
\bibitem{ref23} J.~S.~Levy, A.~Gondarenko, M.~A.~Foster, A.~C.~Turner-Foster, A.~L.~Gaeta, and M.~Lipson, ``CMOS-compatible multiple-wavelength oscillator for on-chip optical interconnects,'' Nat. Photon. \textbf{4}, 37--40 (2010).
\bibitem{ref24} H.~Jung, C.~Xiong, K.~Y.~Fong, X.~Zhang, and H.~X.~Tang, ``Optical frequency comb generation from aluminum nitride microring resonator,'' Opt. Lett. \textbf{38}, 2810--2813 (2013).
\bibitem{ref25} M.~Pu, L.~Ottaviano, E.~Semenova, and K.~Yvind, ``Efficient frequency comb generation in AlGaAs-on-insulator,'' Optica \textbf{3}, 823--826 (2016).
\bibitem{ref26} B.~J.~M.~Hausmann, I.~Bulu, V.~Venkataraman, P.~Deotare, and M.~Lo\u ncar, ``Diamond nonlinear photonics,'' Nat. Photon. \textbf{8}, 369--374 (2014).
\bibitem{ref27} S.~Miller, K.~Luke, Y.~Okawachi, J.~Cardenas, A.~L.~Gaeta, and M.~Lipson, ``On-chip frequency comb generation at visible wavelengths via simultaneous second- and third-order optical nonlinearities,'' Opt. Exp. \textbf{22}, 26517--26525 (2014).
\bibitem{ref28} H.~Jung, R.~Stoll, X.~Guo, D.~Fischer, and H.~X.~Tang, ``Green, red, and IR frequency comb line generation from single IR pump in AlN microring resonator,'' Optica \textbf{1}, 396--399 (2014).
\bibitem{ref29} X.~Xue, F.~Leo, Y.~Xuan, J.~A.~Jaramillo-Villegas, P.-H.~Wang, D.~E.~Leaird, M.~Erkintalo, M.~Qi, and A.~M.~Weiner, ``Second-harmonic assisted four-wave mixing in chip-based microresonator frequency comb generation,'' Light: Sci. Appl. \textbf{6}, e16253 (2017).
\bibitem{ref30} X.~Guo, C.-L.~Zou, H.~Jung, Z.~Gong, A.~Bruch, L.~Jiang, and H.~X.~Tang, ``Efficient visible frequency comb generation via Cherenkov radiation from a Kerr microcomb,'' arXiv:1704.04264 (2017).
\bibitem{ref31} T.~Hansson, F.~Leo, M.~Erkintalo, J.~Anthony, S.~Coen, I.~Ricciardi, M.~D.~Rosa, and S.~Wabnitz, ``Single envelope equation modeling of multi-octave comb arrays in microresonators with quadratic and cubic nonlinearities,'' J. Opt. Soc. Am. B \textbf{33}, 1207--1215 (2016).
\bibitem{ref32} X.~Xue, X.~Zheng, and A.~M.~Weiner, ``Soliton trapping and comb self-referencing in a single microresonator with $\chi^{(2)}$ and $\chi^{(3)}$ nonlinearities,'' Opt. Lett. \textbf{42}, 4147--4150 (2017).
\bibitem{ref33} M.~Karpov, H.~Guo, A.~Kordts, V.~Brasch, M.~H.~P.~Pfeiffer, M.~Zervas, M.~Geiselmann, and T.~J.~Kippenberg, ``Raman Self-Frequency Shift of Dissipative Kerr Solitons in an Optical Microresonator,'' Phys. Rev. Lett. \textbf{116}, 103902 (2016).
\bibitem{ref34} T.~Herr, V.~Brasch, J.~D.~Jost, I.~Mirgorodskiy, G.~Lihachev, M.~L.~Gorodetsky, and T.~J.~Kippenberg, ``Mode Spectrum and Temporal Soliton Formation in Optical Microresonators,'' Phys. Rev. Lett. \textbf{113}, 123901(2014).
\bibitem{ref35} X.~Yi, Q.-F.~Yang, X.~Zhang, K.~Y.~Yang, X.~Li, and K.~Vahala, ``Single-mode dispersive waves and soliton microcomb dynamics,'' Nat. Comm. \textbf{8}, 14869 (2017).
\bibitem{ref36} C.~Bao, Y.~Xuan, D.~E.~Leaird, S.~Wabnitz, M.~Qi, and A.~M.~Weiner, ``Spatial mode-interaction induced single soliton generation in microresonators,'' Optica \textbf{4}, 1011--1015 (2017).
\bibitem{ref37} E.~Obrzud, S.~Lecomte, and T.~Herr, ``Temporal solitons in microresonators driven by optical pulses,'' Nat. Photon. \textbf{11}, 600--607 (2017).
\bibitem{ref38} V.~Ulvila, C.~R.~Phillips, L.~Halonen, and M.~Vainio, ``Frequency comb generation by a continuous-wavepumped optical parametric oscillator based on cascading quadratic nonlinearities,'' Opt. Lett. \textbf{38}, 4281--4284 (2013).
\bibitem{ref39} I.~Ricciardi, S.~Mosca, M.~Parisi, P.~Maddaloni, L.~Santamaria, P.~D.~Natale, and M.~D.~Rosa, ``Frequency comb generation in quadratic nonlinear media,'' Phys. Rev. A \textbf{91}, 063839 (2015).
\bibitem{ref40} X.~Guo, C.-L.~Zou, H.~Jung, and H.~X.~Tang, ``On-Chip Strong Coupling and Efficient Frequency Conversion between Telecom and Visible Optical Modes,'' Phys. Rev. Lett. \textbf{117}, 123902 (2016).
\bibitem{ref41} Y.~Wang, F.~Leo, J.~Fatome, M.~Erkintalo, S.~G.~Murdoch, and S.~Coen, ``Universal mechanism for the binding of temporal cavity solitons,'' Optica \textbf{4}, 855--863 (2017).
\bibitem{ref42} H.~Taheri, A.~B.~Matsko, and L.~Maleki, ``Optical lattice trap for Kerr solitons,'' Eur. Phys. J. D \textbf{71}, 153 (2017).
\end{thebibliography}
\end{document}